\documentclass[aps,twocolumn,showpacs,showkeys]{revtex4}   
\usepackage{epsfig}   
                  
\begin{document} 
   
\title{First-order phase transitions: 
A study through the parallel tempering method}  
  
\author{Carlos E. Fiore} 
%\email{fiore@if.usp.br} 
\affiliation{Departamento de F\'{\i}sica\\   
Universidade Federal do Paran\'a\\   
Caixa Postal 19044\\   
81531 Curitiba, Paran\'a, Brazil}   
\date{\today}   
   
\begin{abstract}  
 We study the applicability of the {\em parallel tempering method} (PT) in the investigation of first-order phase transitions. In this method, replicas of the same system are simulated simultaneously at different temperatures and the configurations of two randomly chosen replicas can occasionally be interchanged. We apply the PT for the Blume-Emery-Griffiths (BEG) model, which displays strong first-order transitions at low temperatures. A precise estimate of coexistence lines is obtained, revealing that the PT may be a successful tool for the characterization of discontinuous transitions.
\end{abstract}  
 
 
  
\maketitle   
  
%%%%%%%%%%%%%%%%%%%%%%%%%%%%%%%%%%%%%%%%%%%%%%%%%%%%%%%%%%%%%%%%%%%%%%%%%%  
\section{Introduction}
Due to the absence of exact  solutions on most systems, Monte Carlo methods play an important role
 not only in statistical physics and critical phenomena but also in other areas.  Usually, 
 the Metropolis \cite{metr} and the Glauber \cite{glauber}  algorithms are used to lead the 
system to the Gibbs distribution. Despite their simplicity and generality,  difficulties 
appears in studying the emergence of phase transitions when they are used to generate the
 microscopic configurations. Several techniques have been proposed to circumvent these difficulties,
 such as the multicanonical technique \cite{berg}, cluster algorithms (that work properly not only 
for reducing critical slowing down  \cite{sw}, but also for eliminating metastability in first-order 
transitions \cite{bouabci,rachadi,fiore6}), the Wang-Landau method \cite{wang},
 simulated tempering \cite{parisi}, and replicas exchanges also named {\em parallel tempering methods} 
(PT) \cite{nemoto,geyer}. 

Special attention has been devoted to this latter approach, due  to its relative simplicity in 
comparison with other approaches and its enormous applicability for several systems in the 
framework  of both 
statistical mechanics \cite{spinglass,proteins,juan,juan1} and molecular dynamics \cite{juan1,hansmann1}. 
Essentially, the PT consists of simulating simultaneously a given set of  replicas of the same 
system  at different temperatures and, occasionally, interchanging the configuration of two
 randomly chosen elements of those replicas. This exchange between pairs of replicas allows 
for the implementation of an ergodic walk in the configuration space when the elements of 
the pair are separated by large free energy barriers.

Although the  PT has been widely used in several contexts, an open question concerns
 its applicability for the investigation of first-order phase transitions \cite{nemoto}.
 In fact, since in discontinuous transitions a gap in the energy 
might lead to a small probability of accepting exchanges between replicas this appears not 
to be a favorable scenario for PT.

 In this paper, we give a further step in this direction by applying the PT to study 
and characterize first-order transitions. We will consider the well known spin-1 
Blume-Emery-Griffiths (BEG) model \cite{BEGMODEL}, which possess a rather rich
 phase diagram with different structures, including first-order transitions in 
the regime of low temperatures. As we shall see, the PT can be 
applied because thermodynamic properties are actually described by continuous
 functions when finite systems are simulated. In fact, the discontinuity of
 thermodynamic properties occur only in the thermodynamic limit. However, 
smooth curves are obtained only when one uses a dynamics yielding a correct 
sampling of the configuration space \cite{bouabci,fiore6,wang}. In particular, the 
use of the PT allows for applying a new finite size procedure for the study 
of first-order phase transitions,  as proposed in Ref. \cite{fiore6}. 
It is worth mentioning that a PT-based analysis of 
first-order transitions has recently been proposed by Neuhaus et al \cite{hansmann2}. 
Such approach is, however, rather different from the one adopted here.

 This paper is organized as follows: In Sec. II we present the model, in Sec. III 
we describe the PT, in Sec. IV we discuss the numerical results, and in Sec. V the conclusions.

\section{Model}
The spin-1 BEG model is  described by the following Hamiltonian:
\begin{equation}
{\cal H}=-J\sum_{(i,j)}\sigma_{i}\sigma_{j}-K\sum_{(i,j)}
\sigma_{i}^{2}\sigma_{j}^{2}+D\sum_{i}\sigma_i^2,     
\label{1} 
\end{equation}
 where $\sigma_i$  denotes the spin variable of the $i$--th site of the lattice 
which assumes the values $\pm 1$ and $0$ 
and the sums run over the nearest neighbor spins on a $d-$dimensional lattice with $V=L^{d}$ sites.
Parameters $J,K$ are the nearest-neighbor spin couplings and $D$ is the quadrupole moment. 
 We have two order parameters defined as follows: 
$q=\langle \sum_{i=1}^{V} \sigma_{i}^{2}\rangle/V$ and $m=\langle \sum_{i=1}^{V} \sigma_{i}\rangle/V$. 
The BEG model will be consider for a square lattice  and  periodic boundary conditions.

\section{PARALLEL TEMPERING  METHOD}
In the parallel tempering method (PT), configurations at high temperatures are used to perform an ergodic 
walk in low temperatures. To this end, we simulate, for fixed values of $D$, a set of  $N$  replicas 
in the  interval of temperatures $\{T_1,...,T_N\}$, where $T_1$ and $T_N$ are  extreme temperatures.

 The dynamics is composed of two parts. In the first part, each one of the $N$ replicas 
are simulated according to the Metropolis algorithm. For the $i$--th replica a given site 
$k$ of the system is chosen at random and we select, with equal probability, one of the two other 
possible spin values $\sigma_{k}'$ such that $\sigma_{k}'\neq\sigma_k$. The spin variable $\sigma_k$ 
is then replaced with $\sigma_k'$ according to  the Metropolis prescription: 
$p_k=\rm min\{1,exp(-\beta \Delta \cal H)\}$ \cite{metr}, where 
$\Delta {\cal H}={\cal H}(\sigma_{k}')-{\cal H}(\sigma_{k})$ and $\beta=1/k_B T$. 
In the second part of the dynamics, the PT is implemented. After a given number
 of Monte Carlo steps, the exchange of configurations of two replicas at the temperatures $T_{i}$ 
and $T_{j}$  are performed with the
 probability $p_{ij}=\min \{1,\exp[(\beta_i-\beta_j)({\cal  H}(\sigma_{i}
)-{\cal H}(\sigma_{j})] \}$, where $T_j > T_i$, $j=i+\delta$, and $\delta$ denotes 
the ``distance'' between two arbitrary replicas. The probability $p_{ij}$ 
depends on $(\beta_{i}-\beta_{j})$ and for this reason the performance of 
method will depend on the ``distance''  between  the replicas. If the 
difference is large enough exchanges will be hardly performed and the 
 PT will not provide any improvement in the results.

In this paper, we adopt two independent procedures to choose the interval of temperatures. 
 In the first  one, the distance between adjacent temperatures obey a geometric 
progression. Some authors have shown \cite{helmut,predescu} that while
this procedure works well when specific heat of the system is about constant, at the emergence 
of a phase transition,  when the specific heat diverges, its efficiency is reduced. 
For this reason, we adopted a second procedure, that consists in distributing temperatures 
in regular intervals between $T_1$ and $T_N$ for a given  small size system. 
By increasing $L$, we {\bf introduce} additional temperatures between $T_i$ and $T_{i+1}$. 
This procedure is necessary because the exchange probability in general decreases as $L$ 
increases.
 We have verified that both procedures lead to the same results, within  of the 
statistical errors.

 Concerning the replicas exchanges we also consider exchanges 
between nonadjacent sites. This is  implemented in this work by allowing 
$\delta$ to range in the interval $\delta=1,..,6$. As it will be shown, although  
non-adjacent exchanges have been less studied \cite{juan,juan1}, because the probability 
of performing a given exchange decreases when $\delta$ increases, they have revealed 
to be essential mechanisms in eliminating hysteretic effects.

\section{Numerical results}
 We have simulated three different values of $K/J$, given by $K/J=$0, 3, and 3.3. 
Note that the first case ($K/J=0$) corresponds to the well known Blume-Capel model. 
Replicas are distributed in the intervals $T_1=1.5 \leq T \leq 2.2=T_N$, for $K/J=3$ 
and $3.3$, and $T_1=0.4 \leq T \leq 0.62=T_N$, for $K/J=0$. We have simulated system with size $L$ 
ranging from $L=10$ up to $L=40$ and we considered $8\times 10^{7}$ Monte Carlo steps to evaluate
 the appropriate quantities after equilibrating the system. For all values of $K/J$ considered here, 
the system displays two ferromagnetic phases (rich at spins $+$ and $-$) for small values of $D$. 
Also, a paramagnetic phase (rich at spins 0) takes place for high values of $D$. A strong
 first-order phase transition between the ferromagnetic and paramagnetic phase occurs for 
a given  value of $D$ that depends on $K/J$ and $T$.
%--------------------- figure -------------------------                        
\begin{figure}
\setlength{\unitlength}{1.0cm}
\includegraphics[scale=0.28]{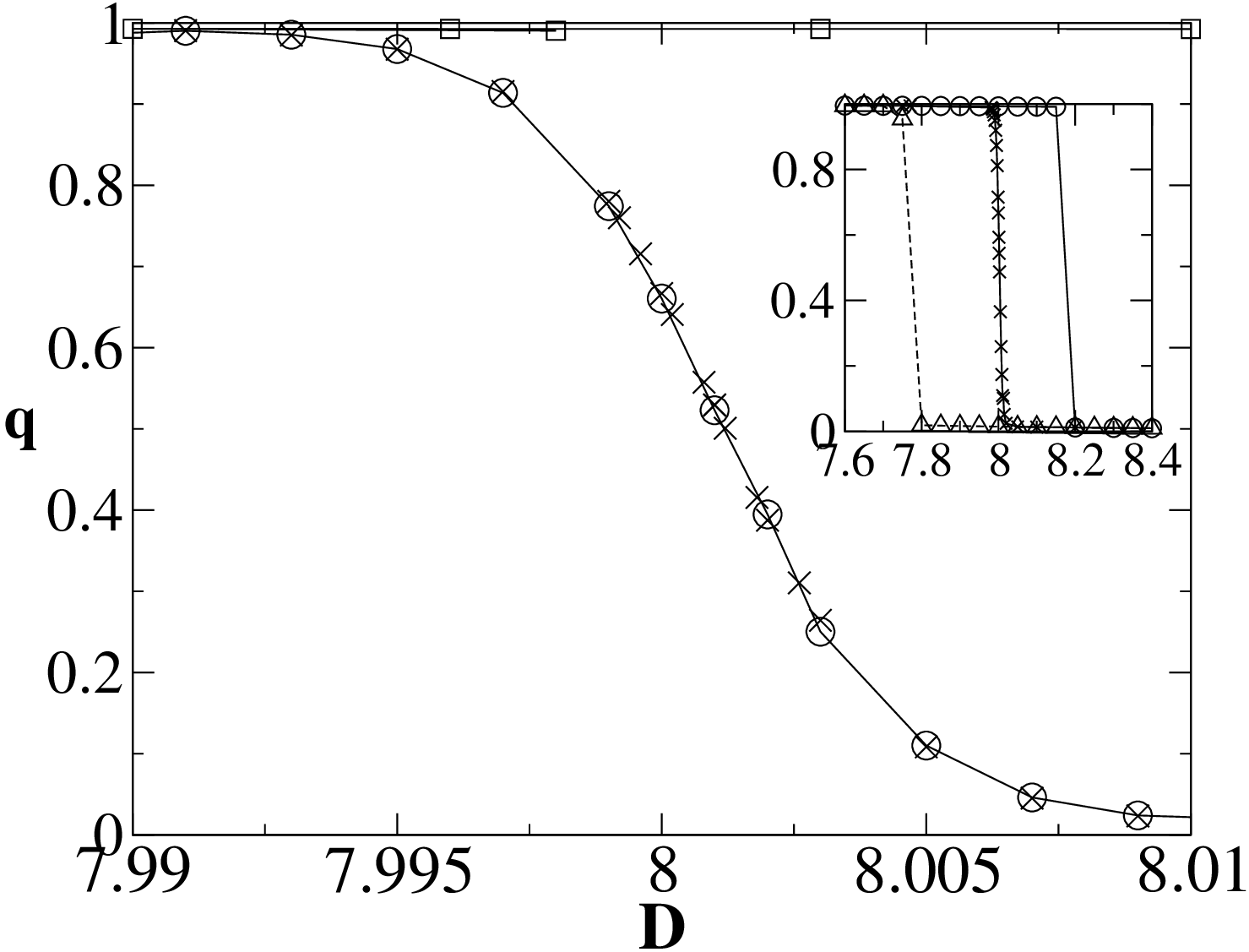}
\caption{Order parameter $q$ as a function of $D$ for $K/J=3$, $T=1.5$
and $L=30$ obtained from parallel tempering (symbol $\times$) and cluster
algorithms (circles). Squares correspond to data 
obtained from parallel tempering with exchanges only between nearest-neighbor
replicas. In the inset,
 circles and triangles refer to the Metropolis algorithm,
whereas the symbol $\times$ refers to the parallel tempering. }
\label{fig1}
\end{figure}
%------------------------------------------------------ 

The first inspection about the applicability of the  PT 
for first-order transitions is shown in the inset of Fig.~\ref{fig1},
 where we compare the PT results with those obtained by 
 using only the Metropolis algorithm. By simulating only with the 
Metropolis algorithm, the system gets trapped in metastable 
states and even after $8\times 10^{7}$ MC steps it does not 
undergo a transition to the stable phase. This effect does not 
occur when we use the PT with nonlocal exchanges, since the system 
becomes able to pass from one phase to the other. The efficiency of the PT
is also corroborated by the agreement with results obtained 
from cluster algorithms \cite{fiore6}, where a smooth curve is 
obtained for the order parameter. However, as it was mentioned 
previously, when one considers only exchanges of configurations 
between nearest-neighbor replicas, hysteresis are still  present, as showed
in Fig.\ref{fig1}. 

\begin{figure}
\setlength{\unitlength}{1.0cm}
\includegraphics[scale=0.36]{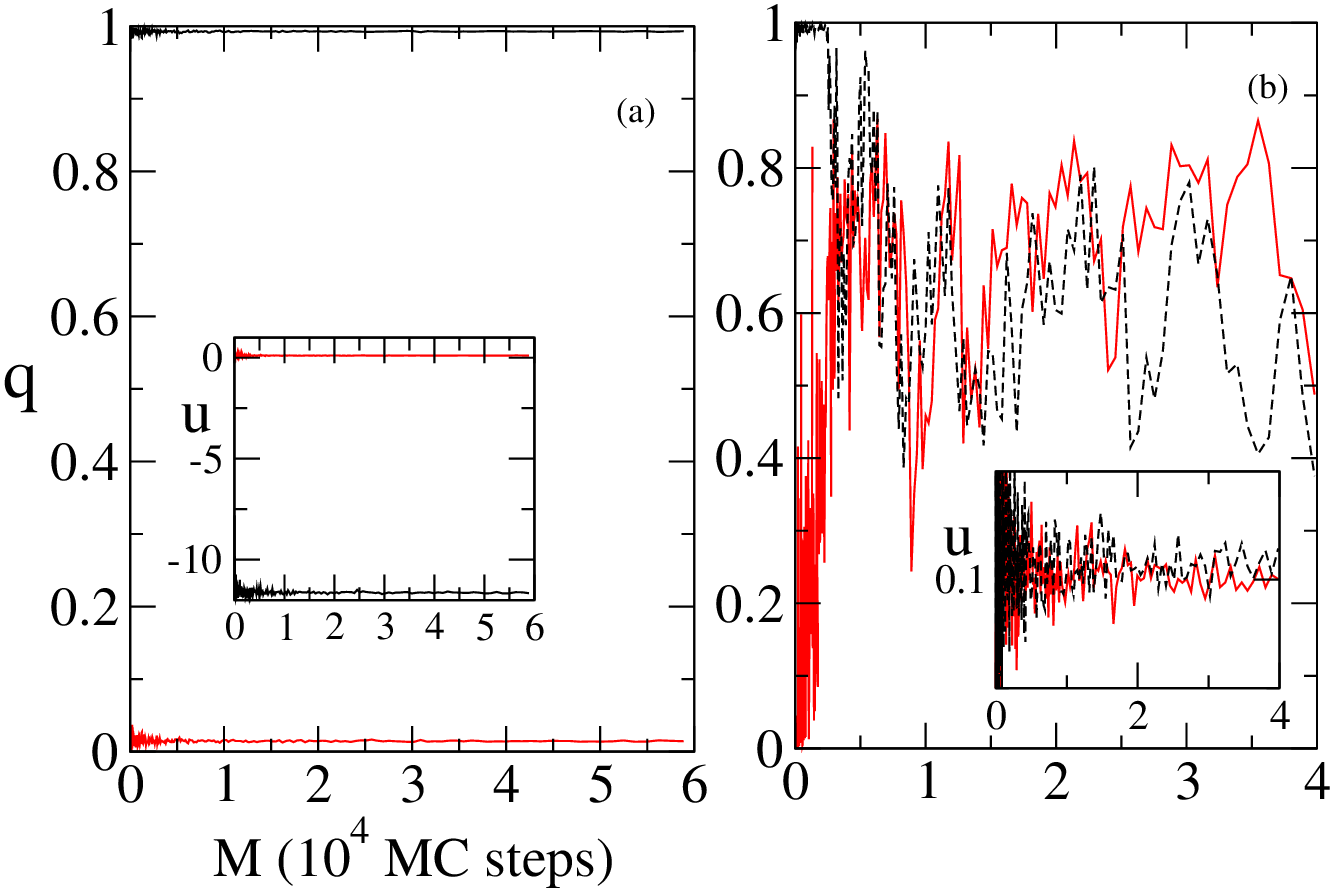}
\caption{Time evolution of  the order parameter $q$ for a single run starting 
from two independent initial configurations  simulated with (a) the Metropolis 
algorithm and $(b)$ the PT, for $L=20$, $T=1.5$, $D=8.0$, and $K/J=3$. In the
  insets the time evolution of the total energy per volume $u$ is 
given for those initial configurations. In contrast with the PT, until $M=6\times 10^{4}$ 
MC steps the Metropolis algorithm provides a nonergodic simulation.}
\label{fig11}
\end{figure}

\begin{figure}
\setlength{\unitlength}{1.0cm}
\includegraphics[scale=0.36]{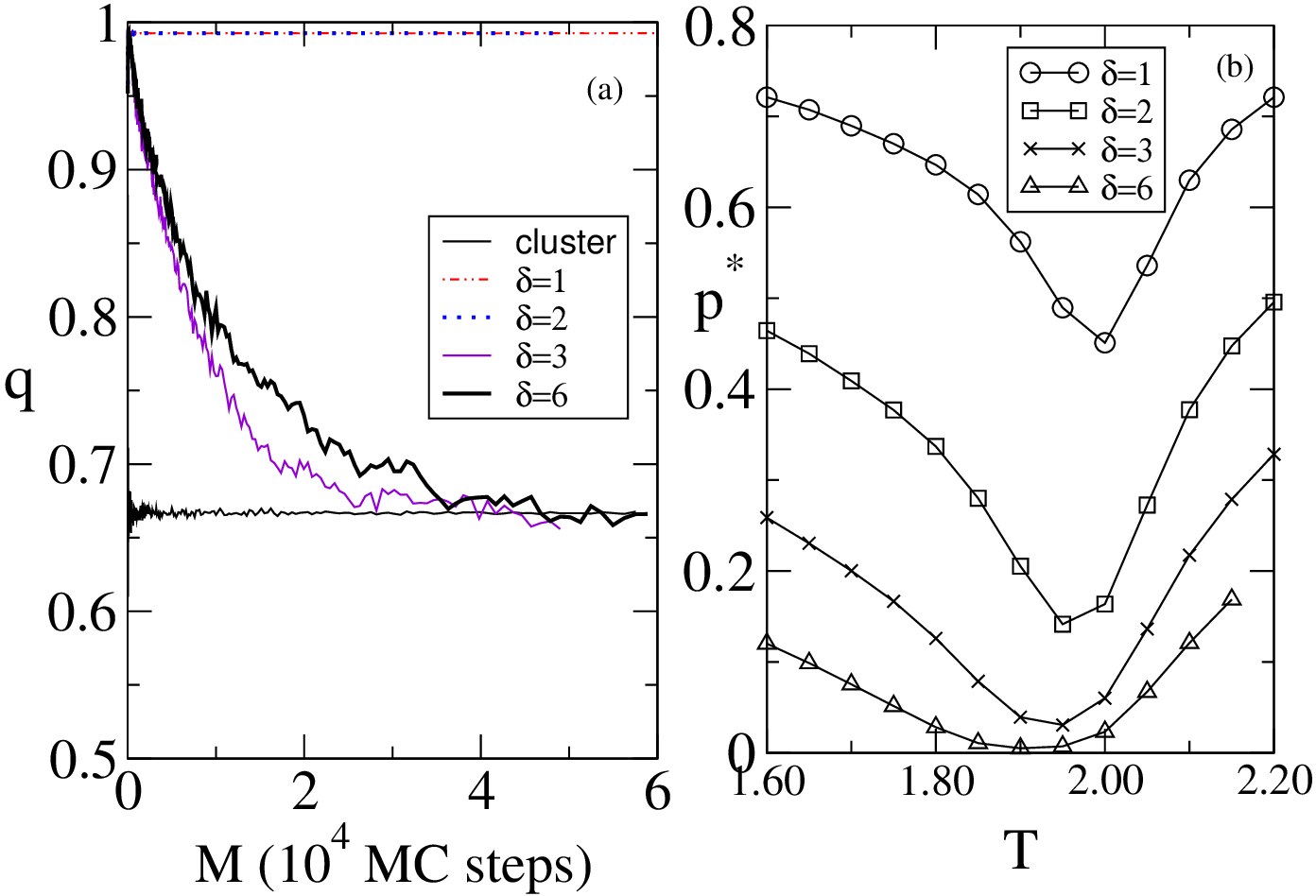}
\caption{In graph $(a)$ we  show the time evolution of the order parameter $q$ 
simulated  by cluster algorithm and  PT with exchanges between $i$ 
and its $i+\delta$th next neighbor replica ($\delta=1,2,3$ and $6$) for $L=20$
 and $NR=500$ independent runs. In graph $(b)$ it is shown the
 mean probability $p^{*}$ versus $T$ for  different  $\delta$.}
\label{fig12}
\end{figure}

The role of non-local exchanges is  analyzed in more details by 
considering the time evolution of  thermodynamic properties at the
 phase coexistence.  In  Fig.~\ref{fig11} we plot, for a single run, 
the order parameter $q$ starting from two different initial configurations 
for $K/J=3$, $T=1.5$ and $L=20$. In the inset of each graph, we  plot
the time evolution of the total energy per volume $u$ for the same initial 
configurations. In contrast  with the  PT, until $M=6\times 10^{4}$ MC steps, 
 the simulation is  not  ergodic when the system is simulated with the Metropolis algorithm.
%--------------------- figure -------------------------                        
%------------------------------------------------------ 
 Next, in  Fig. \ref{fig12} $(a)$ the time evolution of the system simulated 
via PT with local and non-local exchanges is comapared with the results provided 
by cluster algorithms. Note that for $\delta>2$ and $M>3\times 10^{4}$ 
MC steps, the time evolution  the PT simulation for $q$ converges to 
$q\approx 2/3$ (as will be explained later), in agreement with cluster 
algorithm simulations. A similar behavior is obtained in all cases for the quantity $m$. 
In Fig. \ref{fig12} $(b)$ shows the exchange mean probability 
$p^{*}=\langle \min \{1,\exp[(\beta_i-\beta_j)({\cal  H} (\sigma_{i})-{\cal H}(\sigma_{j})] \} 
\rangle$ \cite{juan} as a function of $T$ for different distances $\delta$ 
between replicas and $L=20$. Except for  $\delta=1$, the minimum in $p^{*}$ 
occurs at $T \approx 1.95$, indicating the coexistence between the ferromagnetic
 phases, paramagnetic rich at spins 0 and a disordered phase, that takes place 
in the limit of high temperatures \cite{BEGMODEL}. Our results show that, 
although non-local exchanges are performed less frequently than local ones, 
they are fundamental for ensuring an ergodic simulation of the system. 
%--------------------- figure -------------------------                        
%------------------------------------------------------ 
Next, we will describe the methodology  employed in
determining coexistence lines.  Their location will be derived
from finite size analysis for both the order parameter $q$ and the susceptibility $\chi_T$.

\begin{figure}
\setlength{\unitlength}{1.0cm}
\includegraphics[scale=0.43]{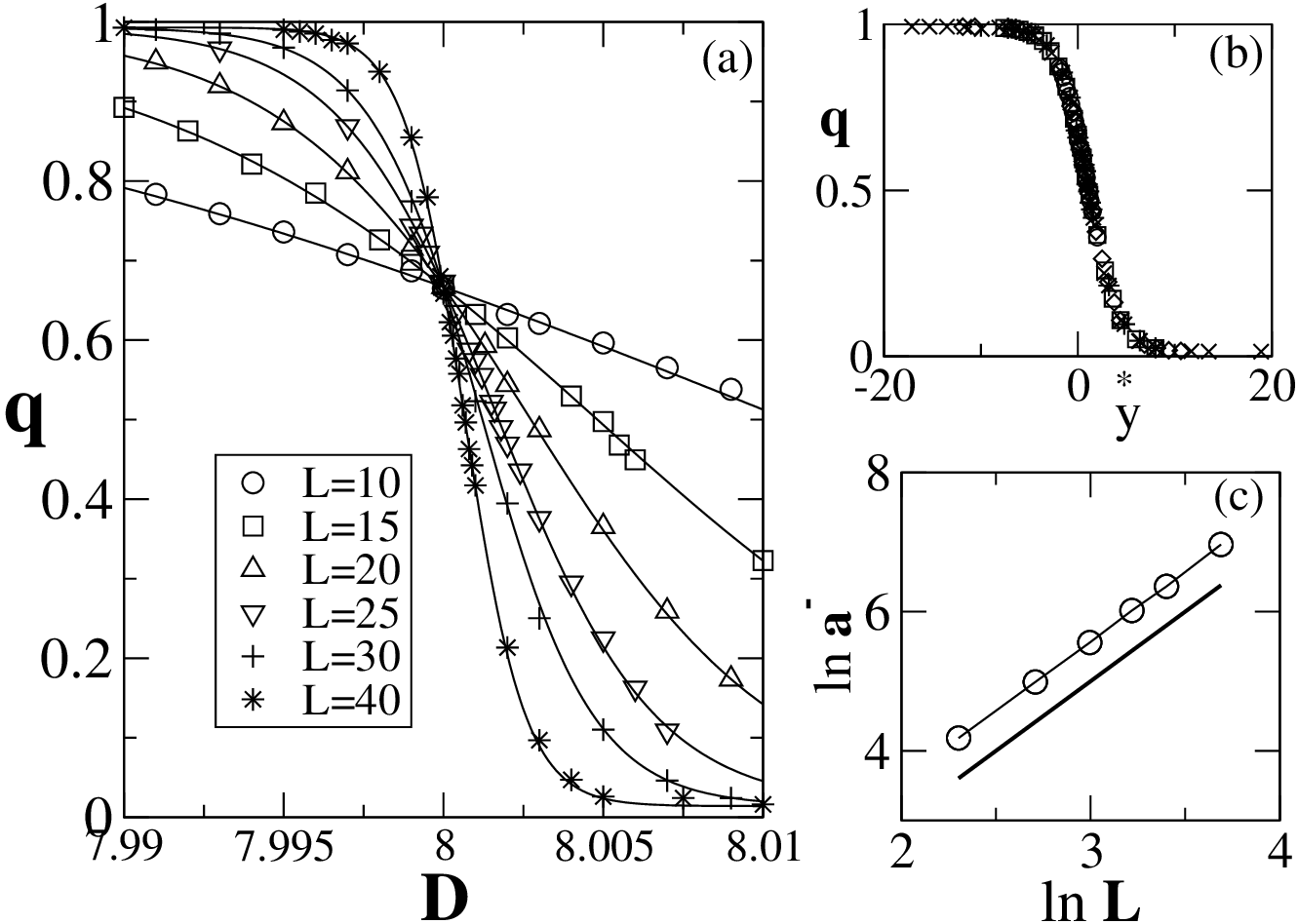}
\caption{Order parameter per volume $q$ versus $D$ for several values
of  the system size $L$ for $K/J=3$ and $T=1.5$. Continuous lines 
 correspond to the fittings defined by Eq. (\ref{e29}). 
In $(b)$ we have a collapse of all data  by using the relation
$y^{*}= (D-D_{0}^{*})L^2$. In $(c)$ we have the log-log plot 
 for the quantity ${\bar a}$ as a function of $L$. 
The straight line has slope $2.00(1)$.}
\label{fig2}
\end{figure}

Although a discontinuous phase transition is characterized by a jump in the order parameter, 
the discontinuity takes place only in the thermodynamic limit. For finite systems, not only 
the order parameter, but also other quantities are described by continuous functions 
\cite{fiore6,wang}.  In such case, the behavior of physical quantities scales with 
the volume of the system \cite{rBoKo,challa}. In Fig. \ref{fig2}, the order parameter 
$q$ is shown as a function of $D$ for
several values of  $L$.
 
Although isotherms present strong dependence on the system size, 
 they intersect one another at the point $D=D^{*}_{0}=8.0000(1)$ 
and $q \approx  2/3$.  As it was explained in Refs. \cite{fiore4,fiore6} 
by means of two different reasonings, the  point where all isotherms cross is
 independent of the lattice size. This  can be understood recalling that in the 
regime of low temperatures two ferromagnetic phases  ($q \approx 1$) coexist with
 a paramagnetic phase rich at spins zero ($q \approx 0$) at $D=D^{*}_{0}$, 
yielding $q \approx 2/3$ for all system sizes. Therefore, the crossing point 
can be used as a criterion to estimate the transition point. 
 As it will be shown later, the estimate of $D^{*}_{0}$ agrees very well 
with the value $D^{*}_{\infty}$ obtained from finite size analysis  for
the susceptibility $\chi_{T}$.  In Fig.~\ref{fig2}(b), we describe the 
collapse of all data by the expression $y^{*}=(D-D^{*}_{0})L^{2}$ confirming  
the dependence on the volume.  At low temperatures, the relation between 
$q$ with the system size $L$ and $D$ is expressed by the following equation \cite{fiore6,fiore5}
\begin{equation}
q = \frac{b+c e^{-{\bar a}z}}{1+d e^{-{\bar a}z}},
\label{e29}                                                                    
\end{equation}                      
where ${\bar a}$, $b$, $c$ and $d$ are 
 fitting parameters and $z \equiv D-D^{*}_{0}$.  
In Fig.~\ref{fig2} (a), continuous lines correspond to 
the  fittings proposed by Eq.~(\ref{e29}). The parameter ${\bar a}$  
scales with the volume, as shown  in Fig. \ref{fig2}(c). 
In the thermodynamic limit $L\rightarrow \infty$, 
 while the quantity ${\bar a}$ diverges the order parameter $q$ does not. 
According to Eq. (\ref{e29}), 
in the ferromagnetic phase, which occurs in the region $D-D_{0}^{*}<0$, 
 we have  that $q\rightarrow c/d$ as $L \rightarrow \infty$. 
 On the other hand, in the paramagnetic phase,  which appears 
for  $D-D_{0}^{*}>0$, $q\rightarrow b $ as $L \rightarrow \infty$. 
For $D=D_{0}^{*}$, we have a jump in $q$, signing a discontinuous phase transition.
%--------------------- figure -------------------------                        
%------------------------------------------------------

\begin{figure}
\setlength{\unitlength}{1.0cm}
\includegraphics[scale=0.4]{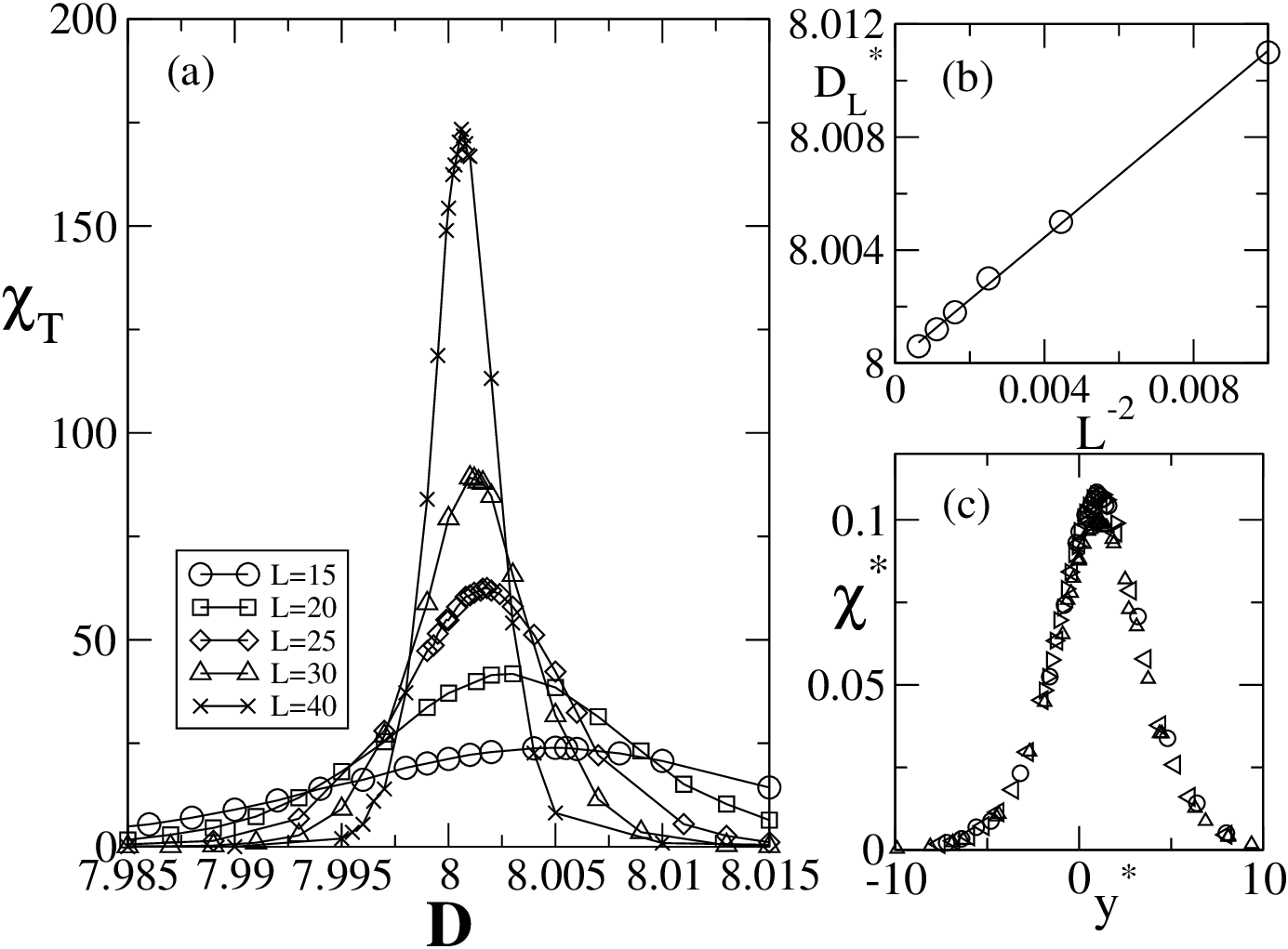}
\caption{Susceptibility $\chi_{T}$ versus $D$  for several values of the system size
 $L$, $K/J=3$, and $T=1.5$. 
In $(b)$, we plot the value of  $D= D_{L}^{*}$
in which $\chi_{T}$ is maximum, as a function of $L^{-2}$. In $(c)$
 we have a collapse of all data using the relations $\chi^{*}=\chi_{T}/L^{2}$ 
and $y^{*}=(D-D_{\infty}^{*})L^{2}$.}
\label{fig3}
\end{figure}

In the second analysis, we determine the transition point by examinating
 the susceptibility $\chi_{T}= \beta L^{2}(\langle q^{2} \rangle-\langle q \rangle^{2})$. 
Increasing $D$ towards the coexistence line, one observes 
{\bf a} sharp peak {\bf in} $\chi_{T}$ at $D^{*}_{L}$ for all system sizes, 
as shown in Fig. \ref{fig3}(a). The deviation between $D^{*}_{L}$ and its 
asymptotic value $D_{\infty}^{*}$ decays as $L^{-2}$ in a first-order 
transition \cite{rBoKo,challa}. Our results satisfy this asymptotic 
relation, as it can be seen in Fig. \ref{fig3} $(b)$. From this law, 
we have obtained the extrapolated value $D_{\infty}^{*}=8.0000(1)$, which agrees 
with the estimate $D_{0}^{*}$ obtained previously and  also agrees with 
the result ${\bar D}=8.0000(1)$, obtained from a cluster algorithm for the
 BEG model \cite{fiore6}. In Fig.~\ref{fig3} $(c)$ we  observe 
that all curves coalesce to $\chi^{*}=\chi_{T}/L^{2}$ and  
$y^{*}=(D_{L}^{*}-D_{\infty}^{*})L^{2}$, confirming once again the scaling with the volume.
%--------------------- figure -------------------------                        
%------------------------------------------------------    

It is worth emphasizing that when one uses only the Metropolis algorithm to 
generate  the configurations, neither the crossing among isotherms nor accurate 
finite size analysis for smooth curves become possible, due to the presence of
hysteresis effects, as it can be seen in Fig. \ref{fig1}.                     

\begin{figure}
\setlength{\unitlength}{1.0cm}
\includegraphics[scale=0.42]{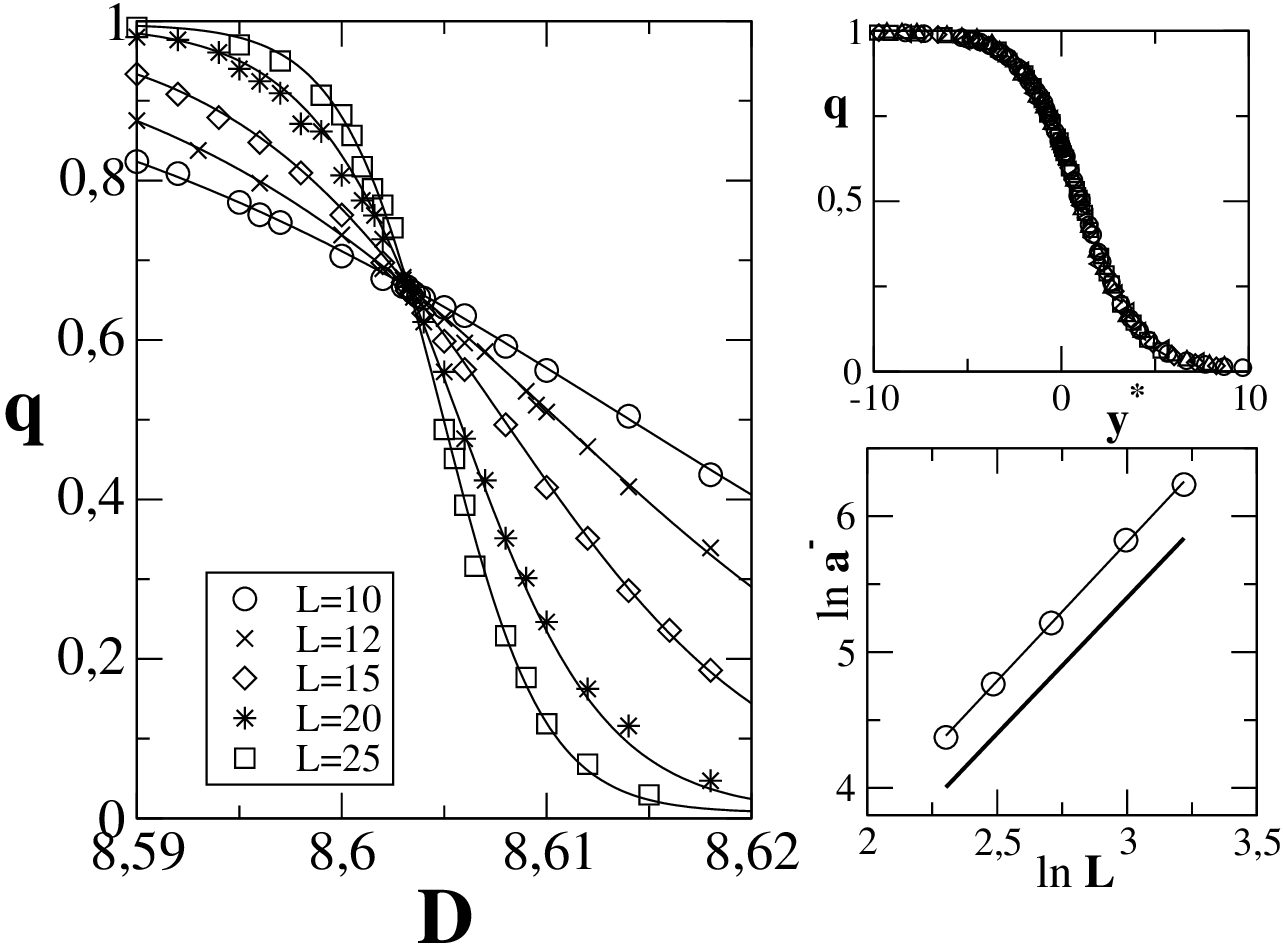}
\caption{Order parameter per volume $q$ 
versus $D$ for several values
of the system size $L$ for $K/J=3.3$ and $T=1.5$. 
Continuous lines  stand for the fittings defined by 
Eq. (\ref{e29}). In $(b)$ we have a collapse of all data
 using the relation $y^{*}=(D-D_{0}^{*})L^{2}$. In $(c)$
 we have  the log-log plot of ${\bar a}$ versus $L$. The straight line has slope $2.00(1)$.}
\label{fig4}
\end{figure}

\begin{figure}
\setlength{\unitlength}{1.0cm}
\includegraphics[scale=0.42]{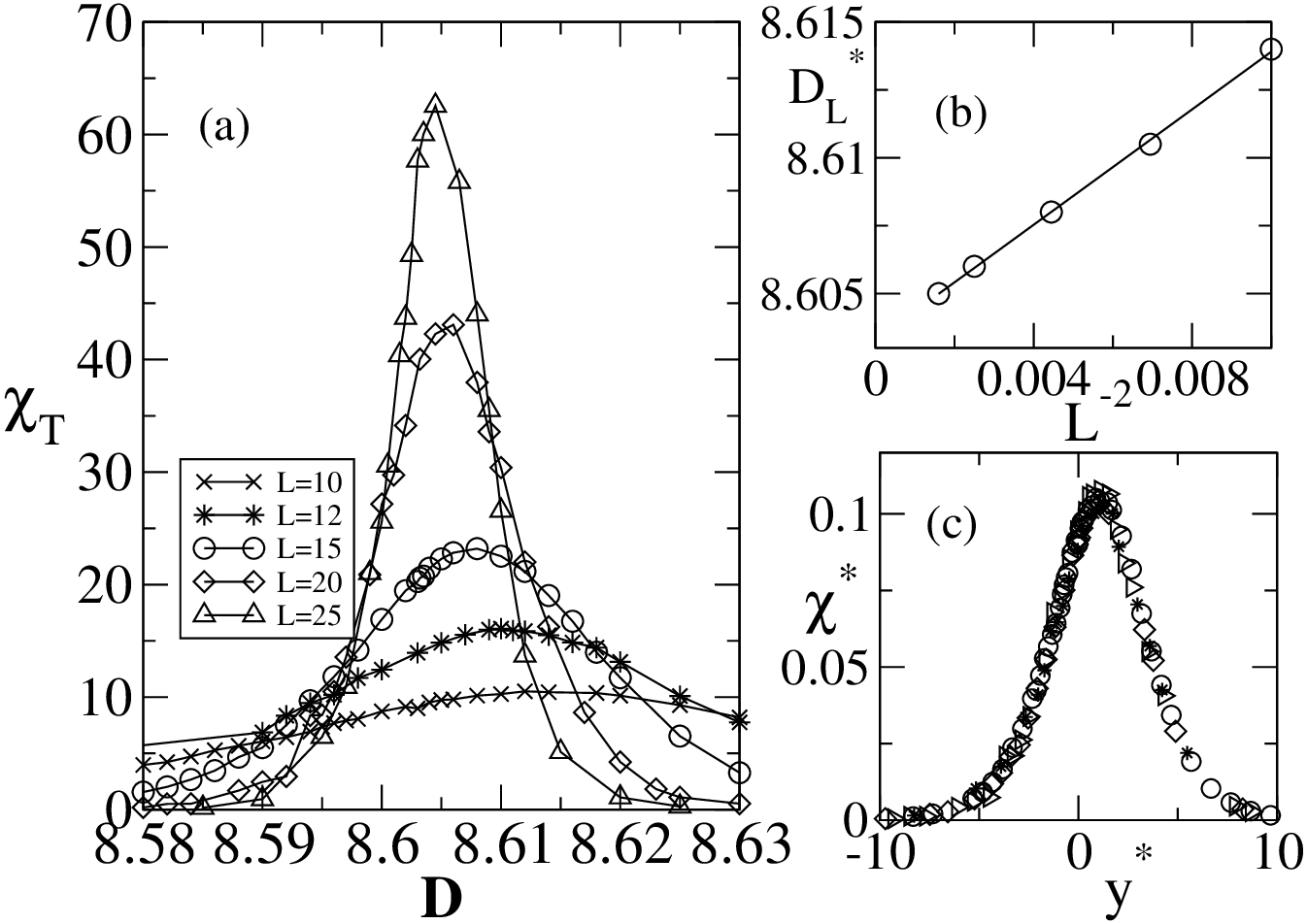}
\caption{Susceptibility $\chi_{T}$ versus $D$ for several values of the system 
size $L$, $K/J=3.3$, and $T=1.5$.  In $(b)$, we plot the value of  $D= D_{L}^{*}$
in which $\chi_{T}$ is maximum, as a function of $L^{-2}$. In $(c)$ we 
have a collapse of all data using the relations $\chi^{*}=\chi_{T}/L^{2}$ 
and $y^{*}=(D-D_{\infty}^{*})L^{2}$. }
\label{fig5}
\end{figure}

In Figs. \ref{fig4} and \ref{fig5}, we repeat, for $K/J=3.3$ and $T=1.5$, both 
 analysis presented above for determining phase coexistence. From the first 
procedure, where all isotherms are  to be fitted by Eq. (\ref{e29}), 
the crossing  is given by $q\approx 2/3$ and $D_{0}^{*}=8.6032(1)$.
 This estimate agrees with the value $D_{\infty}^{*}=8.6033(1)$ 
obtained from finite size analysis for the quantity $\chi_{T}$, as 
showed in Fig. \ref{fig5}. These estimates, both obtained 
by using the  PT, are in good accordance with the 
value ${\bar D}=8.6032$, obtained by Rachadi and  Benyoussef, from  cluster algorithms \cite{rachadi}.
%--------------------- figure -------------------------                        
%------------------------------------------------------                              
%--------------------- figure -------------------------                        
%------------------------------------------------------                        
      
\begin{figure}
\setlength{\unitlength}{1.0cm}
\includegraphics[scale=0.4]{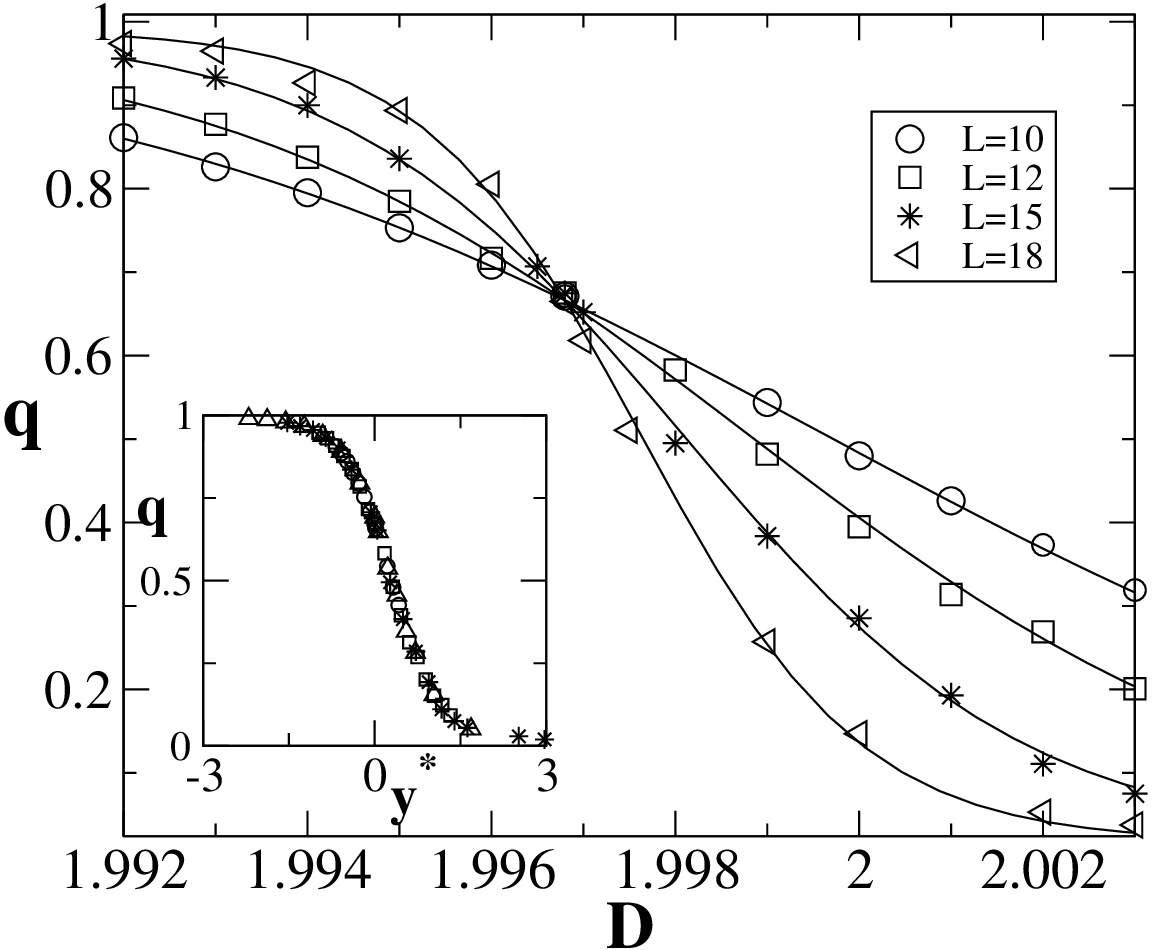}
\caption{Order parameter per volume $q$ versus $D$ for several values  
of system size $L$, for $K/J=0$ and $T=0.4$. Continuous lines  stand for the fittings defined by
 Eq. (\ref{e29}). In the inset, we have a collapse of all data using the relation
 $y^{*}=(D-D_{0}^{*})L^{2}$.}
\label{fig6}
\end{figure}

In the last analysis, we show in Fig.~\ref{fig6} numerical results for $K/J=0$ considering $T=0.4$.
 Fitting all isotherms  with Eq.~(\ref{e29}), the  intersection point turns out to be given
 by $q\approx 2/3$ and $D_{0}^{*}=1.9968(1)$.  The
collapse of data using this estimate of $D_{0}^{*}$ confirms 
again the  adequacy of this procedure for the estimation of transition point. 
Repeating this procedure for $T=0.5$, we verify that all isotherms cross the 
abscissa $D_{0}^{*}=1.9879(1)$, which is in fair agreement with the estimates
$T=0.499(3)$ and ${\bar D}=1.992$, predicted by Wang-Landau method \cite{pla}.
%--------------------- figure -------------------------                        
%--------------------------------
\section{Conclusion}
In this paper, we have applied the parallel 
tempering method (PT) for the study of first-order transitions. 
We have considered different regions of the phase diagram of the BEG model, 
for which usual Metropolis algorithm leads to strong hysteresis at the 
phase coexistence, providing no reliable estimates of the coexistence lines. 
On the other hand, by using the  PT it was possible to circumvent the 
free energy barriers and as a consequence hysteretic effects were eliminated. 
All results obtained  via PT allowed us to locate the transition points 
precisely by means of two techniques, whose estimates agree with those obtained 
from other procedures, such as cluster algorithms  and, in one case, with
the Wang-Landau method. Although the agreement between results obtained from 
PT and  cluster algorithms have been shown to be very well, cluster 
algorithms are more specialized, since each model requires a specific cluster 
algorithm that takes  into account the appropriate transitions. On the 
other hand, PT is general and can be used, in principle, for any 
system. We remark that more
studies of first-order transitions using the parallel tempering
are still required, in order to have a more comprehension of its
performance.
\section{Acknowledgments} 
I acknowledge Juan P. Neirotti, Carlos E. I. Carneiro, Silvio R. Salinas, 
Helmut G. Katzgraber and Renato M. \^Angelo, for a critical reading of this manuscript and
 useful suggestions. This work was partially suported by Funda\c c\~ao de Amparo \`a Pesquisa do Estado 
de S\~ao Paulo (FAPESP) under Grant No. 06/51286-8.  
  
%%%%%%%%%%%%%%%%%%%%%%%%%%%%%%%%%%%%%%%%%%%%%%%%%%%%%%%%%%%%%%%%%%%%%%%%%%  


\begin{thebibliography}{99}
\bibitem{fiore5} 
The expression (\ref{e29}) is derived                                        
from  Borgs and Koteck\`y theory, where it was shown that at low              
temperatures the partition                                                    
function for two or more coexisting phases can be expressed  in terms
of the   
metastable free energy of each  phase. By  taking the
 derivate of
  the logarithm of the partition function                             
with respect to $D$, one gets an expression for                                
$q$, as given in Eq. (\ref{e29}). More details can be found in Ref. 
\cite{fiore6}.
\end{thebibliography}
\end{document}